\documentclass[apj,numberedappendix,revtex4]{emulateapj}
\usepackage{ulem}
\usepackage{epstopdf}
\usepackage{hyperref}
\usepackage{threeparttable}
\usepackage{pdfpages}
\usepackage{csquotes}
\usepackage[multiple]{footmisc}
\usepackage{amsmath}

\def\gtorder{\mathrel{\raise.3ex\hbox{$>$}\mkern-14mu
             \lower0.6ex\hbox{$\sim$}}}
\def\ltorder{\mathrel{\raise.3ex\hbox{$<$}\mkern-14mu
             \lower0.6ex\hbox{$\sim$}}}

\slugcomment{Submitted to ApJ}
\shorttitle{RCF - Redshift Completeness Fraction}
\shortauthors{Kulkarni et al.}

\begin{document}
\title{The Redshift Completeness of Local Galaxy Catalogs}
\author{S.~R.~Kulkarni\altaffilmark{1*}, 
D.~A.~Perley\altaffilmark{2}, \& 
A.~A.~Miller\altaffilmark{3,4}}

\altaffiltext{1}{Division of Physics, Mathematics, and Astronomy, California Institute of Technology, Pasadena, CA 91125, USA}
\altaffiltext{2}{Astrophysics Research Institute, Liverpool John Moores University, IC2 Liverpool Science Park, Liverpool, L3 5RF, UK}
\altaffiltext{3}{Center for Interdisciplinary Exploration and Research in Astrophysics (CIERA) and Department of Physics and Astronomy, Northwestern University, 2145 Sheridan Road, Evanston, IL 60208, USA}
\altaffiltext{4}{The Adler Planetarium, Chicago, IL 60605, USA}
\altaffiltext{*}{E-mail: \texttt{srk@astro.caltech.edu}}

% \email{srk@astro.caltech.edu}
\keywords{galaxies: distances and redshifts --- galaxies: statistics --- supernovae: general}

\begin{abstract}

There is considerable interest in understanding the 
demographics of galaxies within 
the local universe (defined, for our purposes, as the 
volume within a radius of 200\,Mpc or $z\leq 0.05$).
In this pilot paper, using supernovae (SNe) as signposts to galaxies, 
we investigate the redshift completeness of
catalogs of nearby galaxies. In particular, type Ia SNe are 
bright and are good tracers of the bulk of the galaxy population, 
since they arise in both old and young stellar populations.  
Our input sample consists of SNe with redshift $\le 0.05$, 
discovered by the flux-limited ASAS-SN survey.   We define the 
redshift completeness fraction (RCF) as the number of SN 
host galaxies with known redshift prior to SN discovery, 
determined, in this case, via the NASA Extragalactic Database (NED), 
divided by the total number of newly discovered SNe. Using SNe Ia, we find $\mathrm{RCF} = 78 \pm^{6}_{7}$\% (90\% confidence interval) for $z < 0.03$. We examine the distribution of host galaxies
with and without cataloged redshifts as a function of absolute
magnitude and redshift, and, unsurprisingly, find that higher-$z$
and fainter hosts are less likely to have a known redshift prior
to the detection of the SN. However, surprisingly, some
$L_*$ galaxies are also missing. We conclude with thoughts on the future
improvement of RCF measurements that will be made possible from large SN samples resulting from ongoing and especially 
upcoming time-domain surveys.
\end{abstract}

\keywords{galaxies: distances and redshifts --- galaxies: statistics --- SNe: general}

\section{Transients in the Local Universe}
\label{sec:TILU}

Transients in the local Universe provide unique insights into at least
three pressing issues in modern astronomy.  First, nearby events can be studied in great detail, even if their luminosities are relatively low---enabling insights into their physics. A classic example is the detection of SN\,1987A in the Large Magellanic Cloud, which enabled the unambiguous localization of extragalactic neutrinos (e.g., \citealt{mccray93}).
Second, nearby events can be studied demographically to high completeness.  This is important both for obtaining a full understanding of how stars end their lives, and for understanding the role their explosions play in their environments.  For example SNe (SNe), the most commonly observed extragalactic transients, inject energy, momentum and heavy elements into their surroundings.
Relating star-formation rate and chemical abundance to SN
rates is a fundamental exercise in modern astronomy. Nearby (volume limited) SN surveys are needed to provide the latter. 

Third, over the last decade or so, exotic explosive sources have
been identified -- Ultra High Energy Cosmic Rays (UHECR), ultra-high
energy neutrinos, and Gravitational Wave (GW) sources. The horizon
for detecting these sources is limited by either physical phenomena
(the Greisen-Zatespin-Kuzmin effect for UHECRs), or set by the
sensitivity of GW telescopes.\footnote{In this paper, we restrict
the discussion of GW sources to those involving neutron stars since
only for such events are electromagnetic (EM) counterparts expected.}
The latter consideration leads us to a distance limit of $\sim$200\,Mpc
($z\lesssim 0.05$).  As shown by the rich returns from electromagnetic studies of the neutron star coalescence event, GW\,170817 (e.g., \citealt{aaa+17,aaa+17a}), the study of transients in the
local universe is not only of wide importance but also
timely.

The primary motivation for this paper is the last point discussed above, namely the study of electromagnetic counterparts to GW sources. 
For the next few years the typical localization of GW sources will 
be no better than $\sim$50\,$\deg^2$. Naturally, pursuit of such 
large angle localizations will entail a deluge of false positives 
(e.g., \citealt{ksc+16,scs+16}).  As demonstrated by the steps which led to the discovery of the optical counterpart of GW\,170817, a cost-effective 
approach to both minimizing  the background fog of false 
positives and maximizing early identification  is to use on-sky coincidences with catalogs of nearby 
galaxies (e.g., \citealt{gck+16}).

Additionally, we note that catalogs of nearby galaxies have other uses. For instance,
there is considerable interest in studying the youngest SNe.  SNe take time to brighten to peak luminosity and are very faint in the first hours to days after the initial explosion, so the appearance of a new source with an inferred luminosity much lower than a classical SN at peak is cause for vigorous pursuit.  But determining this luminosity requires knowledge of the redshift, which will be known in advance only if the transient coincides with a galaxy with a cataloged redshift.

As a result of deep wide-field imaging surveys, such as PanSTARRS-1
(PS1) and the Sloan Digital Sky Survey (SDSS), all galaxies,
particularly in the Northern sky, to $\sim$23\,mag, are ``cataloged"
(in the sense that multi-band photometric measurements exist, and
the galaxies are assigned a nominal name).\footnote{Separating stars from galaxies in these photometric catalogs presents a significant challenge, however.}
However, what matters
for the purpose of this paper is a \textit{reliable} galaxy redshift
measurement, \textit{prior} to any transient follow-up (such as during a search for EM counterparts).  The fraction of nearby galaxies of a certain type (for example, candidate hosts of LIGO GW sources) with a redshift that is recorded in published catalogs can be defined as the redshift ``completeness.''\footnote{We will more precisely state our operational definition of ``completeness'' in Section~\ref{sec:Assessing}.}

In this paper, we explore the use of SNe to assess the completeness
of nearby galaxy catalogs.  SNe are very luminous, relatively common (in a cosmological sense), and are found routinely in surveys that now cover the entire sky every few days (and therefore the entire local volume out to some distance limit, subject to the limitation of extinction from the Galactic plane).  They thus provide an effective way of randomly sampling galaxies that is not strongly dependent on the observational properties of those galaxies (in particular, on galaxy luminosity).  Recently, \citet{hbs+17} remarked that 24\% of nearby bright SNe were discovered in cataloged host galaxies without secure redshifts, suggesting that redshift incompleteness may be significant even today.  As a next step, we refine their estimate by formally restricting their sample to a limited volume, and examine in detail those SNe that occur in galaxies whose redshifts have been ``missed'' by spectroscopic surveys.

\section{Catalogs of nearby galaxies}

The construction of redshift catalogs for meaningful numbers of
nearby galaxies can only be accomplished via large-scale spectroscopic
surveys (e.g.\  \citealt{bm09}).  However, given the more than
eight magnitude spread in galaxy luminosities \citep{bm09}, large
numbers of even very nearby galaxies are likely to be quite faint, so
these catalog(s) will necessarily be incomplete.  Despite this challenge,
astronomers have assembled a number of all-sky galaxy catalogs: the 11-Mpc
\textit{Nearby Galaxy Catalog} \citep{kmk13}, \textit{The Extragalactic
Database} \citep{trs+09} and \textit{Hyperleda} \citep{ppt+03}.
These catalogs are linked to the NASA/IPAC Extragalactic Database
(NED).\footnote{\url{https://ned.ipac.caltech.edu/}}
Figure~1 of \cite{gck+16} provides a graphical illustration of the bias of inputs to spectroscopic surveys.

How complete are these catalogs?  One approach is to extrapolate the 
findings of higher-$z$ surveys (which scrutinize small areas of the sky
down to extremely deep limits) down to $z\approx 0$. However, as shown by deviations of
galaxy recession velocities from the Hubble flow \citep{jrk07} and
via direct galaxy catalogs, the local Universe is lumpy (10\%
fluctuations or more) on length scales as large as
200\,Mpc.\footnote{This is not at all surprising given the
150\,Mpc Baryon Acoustic Oscillations (BAO) length scale.}
Extrapolating a local catalog believed to be highly complete (e.g., \textit{Nearby Galaxy Catalog}, which is complete to $M_B<-15$ mag within 11 Mpc outside the Galactic plane) from the bottom up is even more problematic given strong fluctuations on these distance scales (e.g., not a single galaxy cluster lies within 11 Mpc). 

\section{Assessing Catalog completeness with SNe}
\label{sec:Assessing}

Here, we explore the use of SNe to evaluate the completeness of
catalog(s) of the nearby Universe. Type Ia SNe are well suited for
this task. These SNe are luminous (only $\sim$1\,mag fainter than a local
$L_\ast$ galaxy in the $V$ band) and thus easily identified. The
luminosity function is narrow: in $B$ band the luminosity function
can be fit with a Gaussian with a mean value of $-19.4\,$mag and an rms of 0.14--0.23\,mag
\citep{yf10}. Next, SNe Ia arise from both old and young populations \citep{sb05,slp+06}. Thus, they sample all types of galaxies and can be 
used to study galaxy demographics without 
missing significant subpopulations. 

In contrast, core-collapse SNe (CC SNe) arise only in
star-forming galaxies and exhibit a wide luminosity function, with absolute magnitude (SDSS $r$ band) ranging from
$-18$ to $-12$ \citep{tcd+14}. Clearly, additional care is
necessary in using CC SNe to measure the RCF.

Our proposal to assess the completeness of nearby galaxies
catalogs is simple: (1) Following the detection of a SN, measure the
redshift of the host galaxy, $z_\mathrm{host}$. (2) Retain those
events with $z_\mathrm{host} \le z_*$, where $z_*$ is the redshift
to which the completeness is being measured (``tries"). (3) Check
the galaxy catalog(s) to see if the SN host galaxy has a reliable
redshift entry. If present  and if $z_{\rm host}$ is equal to $z_{\rm SN}$ to within, say $0.001$, then it is a ``hit.'' The redshift
completeness factor (RCF) of the galaxy catalog is given by the ratio of
``hits" to ``tries.'' 

We emphasize that our definition of the RCF is explicitly tied to SNe and is not the fraction of galaxies with known redshifts \textit{by number}.  The majority of galaxies within any volume are very small galaxies which are also the most difficult to detect (indeed, new satellites of the Milky Way are still being uncovered). Previous studies \citep{sb05,slp+06} have found that the Ia propensity (the rate of Ia production within a galaxy), $\mathcal{P}$, 
is well represented by a linear combination of the stellar mass of the galaxy ($\mathcal{M}$) and the star formation rate ($\mathcal{S}$).  Thus the RCF (as defined here) approximately measures the completeness weighted by $\mathcal{P}$.  Other definitions could be used: for example, had we employed CC SNe instead of SNe Ia, the resulting RCF would measure the completeness weighted by $\mathcal{S}$ alone.

SN-based estimates of the RCF will depend on the completeness of the parent SN survey(s).  A SN survey that is not complete to the limit of the search volume will find a larger fraction of low-luminosity events at smaller distances whereas more luminous events can be found to larger distances (with concomitant larger volume). Thus, poor control of completeness in a SN survey will bias the sample to galaxies which host more luminous events.  In the case of Ia SNe, a correlation between the luminosity of the SN and the properties of its host does exist \citep{sch+10}, although it is relatively weak.  A bigger concern is if the properties of the host directly affect completeness---for example, if SNe are systematically missed in regions of high galaxy surface brightness, or if the appearance of the host galaxy is a consideration in decisions about spectroscopic classification.

We emphasize that the SN rate per galaxy is far too low to provide a practical means of actually discovering new galaxies in significant numbers.  Instead, the approach advocated here provides a check on the completeness of \emph{existing} catalogs, independent of the traditional luminosity function approach.

\section{Primary Data}
\label{sec:PrimaryData}

\begin{deluxetable}{lrrr}[ht]
\tabletypesize{\scriptsize}
\tablecaption{SN Demographics from the two ASAS-SN surveys}
\tablehead{
\colhead{Catalog} & \colhead{type$^a$} & \colhead{$n$(SN)} & \colhead{$n({\rm NED}_z$)} }
\startdata
\textit{A1} & Ia    & 66 &  51\cr
\textit{A1} & CC  & 25 & 22\cr
\textit{A2} & Ia    & 141 & 101 \cr
\textit{A2} & CC  &  39  & 27   
\enddata
\tablecomments{$n(\mathrm{SN})$ is the number of SNe in each
category. $n({\rm NED}_z)$ is the number of host galaxies 
with a redshift entry in NED (and obtained prior to the SN discovery).} 
\tablenotetext{a}{The type Ia encompasses normal Ia and all sub-types
such as 91T, 91bg, CSM and 00cx.  All non-Ia SN are called
as core collapse (II, IIn, Ibc).} 
\label{tab:SNDemographics}
\end{deluxetable}

To assess the RCF we use NED as our input host galaxy catalog and the first two SNe catalogs (hereafter, {\it A1, A2}) published by the All-Sky
Automated Survey for SN (ASAS-SN) project 
(\citealt{hsk+17, hbs+17}; hereafter, H1 \& H2, respectively). ASAS-SN is well-suited for our purposes.  It is a flux-limited survey, ${\rm V_{\rm peak}\le
17}$\,mag, that covers the entire sky, and is not targeted to specific galaxies.  Additionally, because of its shallow flux limit, ASAS-SN
candidates are bright enough for worldwide follow up using small
telescopes.  As a result the classification (SN type and
$z_\mathrm{host}$) is essentially complete.
In contrast, amateur discoveries (as well as professional surveys such as LOSS, the Lick Observatory SN Survey; \citealt{lft+00}) target well-resolved and bright galaxies and are therefore biased. 
Other, recent untargeted surveys (e.g.\ ATLAS, iPTF, Gaia) are not likely to be strongly biased in terms of discovery,  but the degree of bias in terms of selecting candidates to follow-up (and classify as SNe) is not well-quantified.

Table~\ref{tab:SNDemographics} provides a top level summary of the
two ASAS-SN surveys.
Catalog \textit{A1} lists 91 SNe discovered during the period
2013--2014 (H1). Catalog \textit{A2}
lists 180 SNe found in the calendar year 2015. In \textit{A2} 11
SNe are marginally fainter than 17\,mag (H2). The catalogs also include host galaxy data: name,
redshift, SDSS, \textit{GALEX}, 2MASS and \textit{WISE} photometry.

H1 and H2 do not specifically
identify host galaxies that lack a cataloged redshift prior to the
discovery of the SN. To remedy this situation we wrote a program
to query NED and obtained the redshifts of the putative host
galaxies. We refer to the sample of galaxies with a redshift entry
in NED as the ``NED$_z$" sample and those which lacked an entry as the
``!NED$_z$" sample.

Nine entries (ASAS-SN-15de, 15ji, 15jm, 15lh, 15nh, 15og, 15ts,
15ua, 14ms) have $z_\mathrm{host}>z_\ast = 0.05$. These were
deleted from further consideration. Next, we inspected the
difference between the SN redshift, $z_\mathrm{SN}$, and
the purported host redshift given in NED, $z_\mathrm{host}$.
Bearing in mind the lower precision of the SN redshift we made an allowance
and inspected events with $\vert
z_\mathrm{host}-z_\mathrm{SN}\vert > 10^{-3}$.  The rationale for this choice is that the redshift of the host is usually the systemic velocity of the galaxy (a fiber of the spectrograph
encompasses the central region of the galaxy) whereas the SN will have additional velocity arising from the rotation curve.  
Three events stood
out: ASASSN-15jo has $z_\mathrm{host}=0.014$ (Abell\,S0753),  but
$z_\mathrm{SN}=0.011$; ASASSN-13an has $z_\mathrm{host}=0.02431$ (2MASX J13453653$-$0719350),
but $z_\mathrm{SN}=0.0216$; and ASASSN-15ic has
$z_\mathrm{host}=0.0637$ (2MASX\,J06145320$-$4247357), $z_\mathrm{SN}=0.025$. We retain the
first two events and delete the last one from further
consideration.  
The final sample for the analysis reported below
consists of 261 SNe.

\section{Analysis}
\label{sec:Analysis}

\begin{figure}[htbp] 
   \centering
   \includegraphics[width=3in]{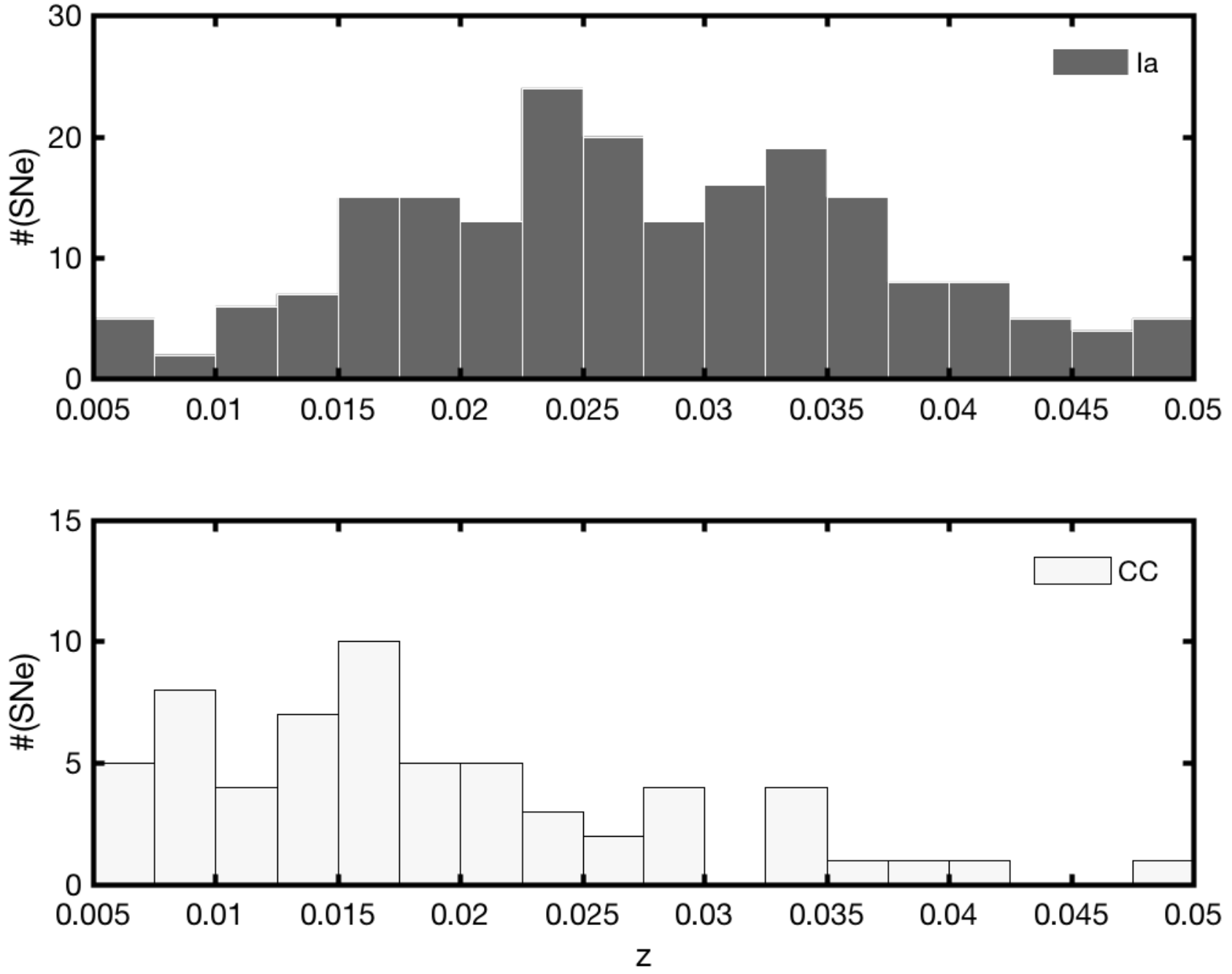}
   \caption{\small
   Histogram of $z\le 0.05$ SNe of type Ia (\textit{top}) and 
   core-collapse (\textit{bottom}) from the \textit{A1} and \textit{A2} catalogs.
 }
\label{fig:RedShiftHistogram} 
\end{figure}

The redshift distribution of ASAS-SN events is displayed in
Figure~\ref{fig:RedShiftHistogram}. SNe Ia peak at $z\approx
0.025$ whereas CC events peak at lower redshift and exhibit
a long tail (as expected, given their lower average luminosities and broad luminosity function). The ``completeness'' of the SN sample itself 
is not the primary topic of this paper.\footnote{Understanding the
completeness of a SN survey is a major project in itself!} What is
central to this paper is that the SN sample not be biased by host
galaxies. So we will proceed with the analysis of the sample we
have in hand. The RCF, assuming NED as the input
host galaxy catalog, for $z_\mathrm{host}\lesssim 0.03$
is $80 \pm 5$\% (90\% confidence interval; see Table~\ref{tab:RCF03}). The precision of these estimates is limited by small number binomial statistics. The RCF as traced by SNe Ia, furthermore, is somewhat lower ($78 \pm^{6}_{7}$\%) than the RCF traced by CC SNe ($87 \pm ^5_{9}$\%), suggesting that galaxy catalogs are more complete for star-forming hosts. The precision of these
estimates is limited by small number binomial statistics. For our full sample ($z_\mathrm{host}\lesssim 0.05$), there are 64
!NED$_z$ galaxies, which corresponds to $\mathrm{RCF} = 75 \pm ^5_{4}$\%. 

\begin{deluxetable}{lllll}[ht]
\tabletypesize{\scriptsize}
\tablecaption{Redshift completeness factor of ASAS-SN SNe ($z<0.03$)}
\tablehead{
\colhead{SN} & \colhead{$n$(SN)\tablenotemark{a}} & \colhead{$n({\rm NED}_z)$\tablenotemark{b}} & \colhead{$n({\rm !NED}_z)$\tablenotemark{c}} &  \colhead{RCF\tablenotemark{d}}}
\startdata
All & 173 & 139 & 34 & 0.75--0.85\cr
Ia & 120 & 93 & 27 & 0.71--0.83\cr
CC & 53 & 46 & 7 & 0.78--0.93
\enddata
\tablenotetext{a}{The total number of SNe.}
\tablenotetext{b}{The number of SNe with a putative host galaxy with redshift listed in NED, prior to SN discovery.}
\tablenotetext{c}{The number of SNe with a putative host galaxy but whose redshift is not listed in NED.}
\tablenotetext{d}{The Redshift Completeness Factor (RCF; see \S\ref{sec:Assessing}) range covers a confidence range of 5\% to 95\% and is obtained
by assuming a flat Bayesian prior. }
\label{tab:RCF03}
\end{deluxetable}

\section{Analysis of Host Galaxy Luminosities}
\label{sec:MissingGalaxies}

The virtue of the SN RCF is that it is a well-defined and simple metric, with no free parameters other than the distance (volume) limit employed.
However, within this volume, we generally expect the completeness to be higher for luminous galaxies (with high stellar mass $\mathcal{M}$,  or high star-formation rate $\mathcal{S}$) than less luminous ones.  Knowing these parameters for the galaxies within our sample, we can subdivide our targets by $\mathcal{M}$ or $\mathcal{S}$ to answer interesting questions such as, ``How complete is our understanding of $\mathcal{M}$ / $\mathcal{S}$ within the local
volume"?   For this purpose, we use the 2MASS $K_s$-band as a proxy for
$\mathcal{M}$ and the \textit{GALEX} NUV band for $\mathcal{S}$, and examine the completeness as a function of these two parameters.

H1 \& H2 took $K_s$-band magnitudes from
2MASS when available; otherwise, a $K_s$-band mag was estimated from \textit{WISE} $W1$ when detected in that band (offset
by a typical $ K_s-W_1=-0.64$\,mag), or (in
the absence of both 2MASS and \textit{WISE}) a limit of $K_s>15.6$\,mag was set.  We employ the same approach using their magnitude catalogs, except that for the last group (3 galaxies in catalog \textit{A1} and 5 in catalog \textit{A2}) we simply set $K_s=15.6$\,mag.\footnote{We note that catalog magnitudes may sometimes significantly underestimate the total flux of extended sources due to the presence of significant flux outside the automatically-defined aperture (Jarrett et al., in prep).  We neglect this effect in our pilot study, but note that any detailed characterization of the host population probed by nearby SNe would require careful attention to this issue.}

In \textit{A1} there are 77 SNe with NUV
data, 5 with only SDSS $u^\prime$ data and 9 with neither NUV nor
$u^\prime$ data. In \textit{A2}, the corresponding
numbers are 138, 17 and 24, respectively. We
took the sample of SNe with both NUV and $u^{\prime}$ data,
applied the Galactic extinction correction, and found the median
of NUV$-u^{\prime}\approx 1\,$mag. We use the NUV data when
available and $u^\prime$ with the aforementioned offset applied
otherwise. We call this hybrid magnitude the ``UV" mag. The 32 SNe
with neither NUV nor $u^\prime$ data are not included in the
analysis.

For the discussion below we note that 
the relevant Schechter characteristic
luminosity parameters are $M^*_{\rm K_s}=-24.2\pm 0.03$
\citep{cnb+01} and $M^*_{\rm NUV}=-18.23\pm 0.11$ \citep{wtm+05}.
Below we examine the missing galaxies with luminosity $L>L_*$
in both the $K_s$ and ``UV'' bands. We then isolate the sample to
just Type Ia SNe to investigate missing galaxies with $L < L_*$.

\begin{figure*}[htp] 
  \centering
  \includegraphics[width=5.5in]{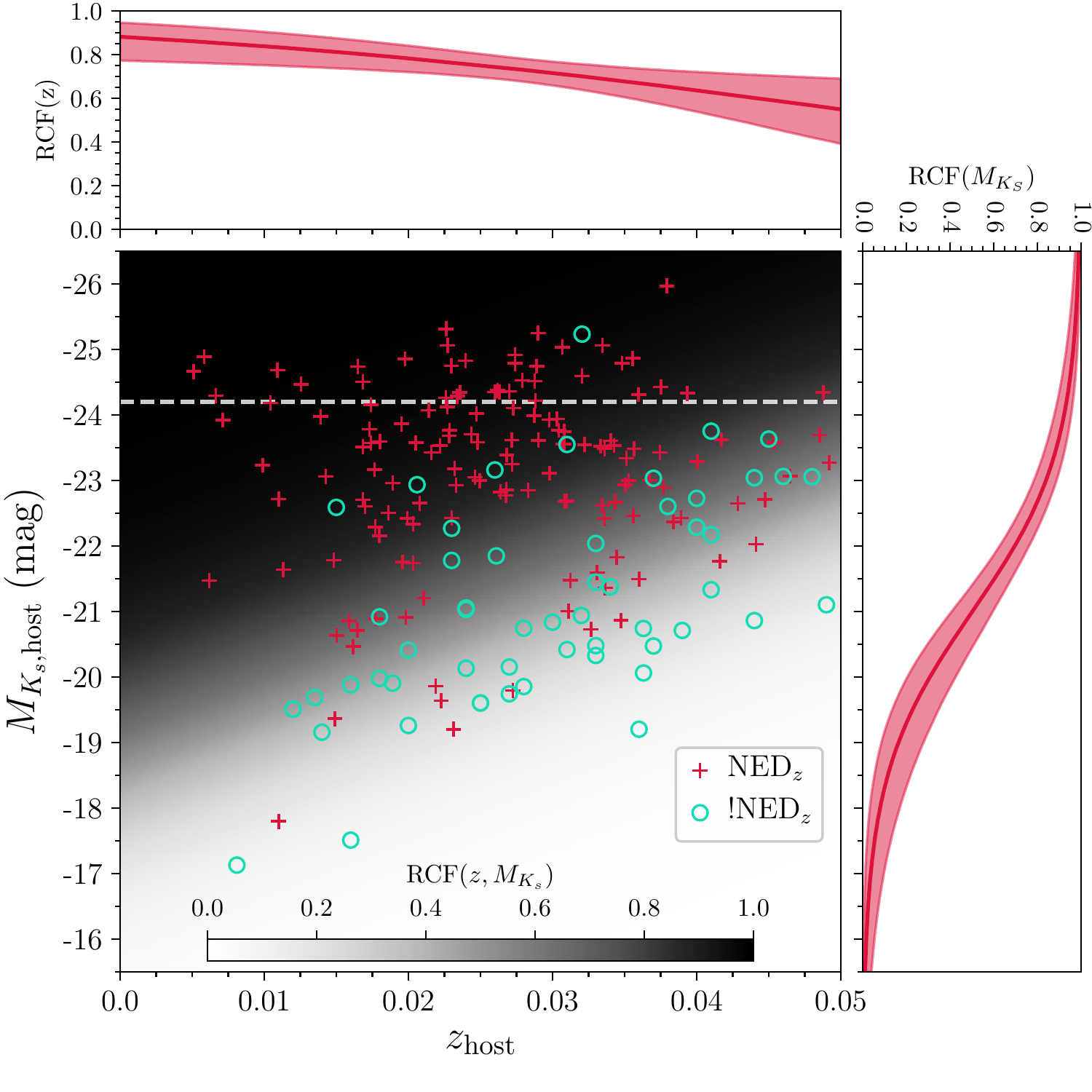} 
  \caption{Absolute $K_s$-band magnitude, $M_{K_s,\mathrm{host}}$, vs.\ redshift, $z_\mathrm{host}$, for the host galaxies of SNe Ia in \textit{A1} and \textit{A2}. Galaxies with redshift entries in NED are shown as red pluses, while those lacking redshifts (!NED$_z$) are shown as teal circles. The horizontal dashed line shows $M_{K_s}^\ast$. The shaded background shows the probability of a host galaxy having a cataloged redshift given its redshift and $M_{K_s}$ ($\mathrm{RCF}(z, M_{K_s})$; see Appendix~\ref{sec:bayes} for details). The top and right plots show the probability of a host galaxy having a cataloged redshift given \textit{only} its redshift, $\mathrm{RCF}(z)$, \textit{or} $M_{K_s}$, $\mathrm{RCF}(M_{K_s})$, respectively. In these two plots the solid lines show the median value of the RCF, while the shaded area corresponds to the 90\% bound on the RCF.
  }
\label{fig:zKmag_Ia}
\end{figure*}

\begin{figure*}[htp] 
  \centering
  \includegraphics[width=5.5in]{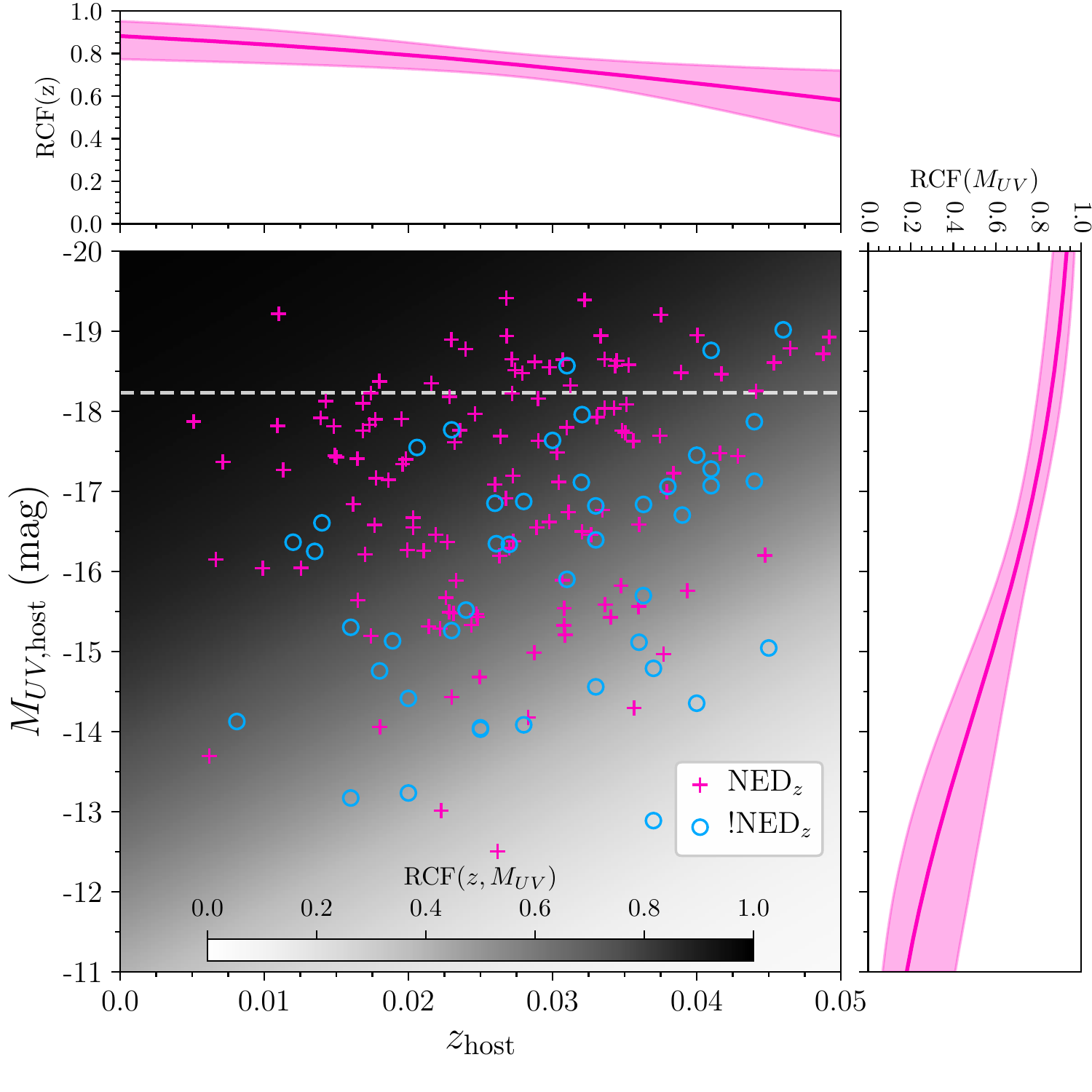} 
  \caption{\small 
  Absolute ``UV''-band magnitude, $M_{UV,\mathrm{host}}$, vs.\
  redshift, $z_\mathrm{host}$, for the host galaxies of SNe Ia in
  \textit{A1} and \textit{A2}. Galaxies with redshift entries in
  NED are shown as magenta pluses, while those lacking
  redshifts (!NED$_z$) are shown as blue circles. 
  The horizontal dashed line shows $M_{UV}^\ast$. The shaded background shows the probability of a host galaxy having a cataloged redshift given its redshift and $M_{UV}$ ($\mathrm{RCF}(z, M_{UV})$; see Appendix~\ref{sec:bayes} for details). The top and right plots show the probability of a host galaxy having a cataloged redshift given \textit{only} its redshift, $\mathrm{RCF}(z)$, \textit{or} $M_{UV}$, $\mathrm{RCF}(M_{UV})$, respectively. In these two plots the solid lines show the median value of the RCF, while the shaded area corresponds to the 90\% bound on the RCF.
  }
\label{fig:zUVmag_Ia}
\end{figure*}

\subsection{$K_s$-band}
\label{sec:Kband}

The absolute $K_s$-band magnitude, $M_{K_s,\mathrm{host}}$, and
redshift, $z_\mathrm{host}$, of each SN Ia host galaxy in our sample
is displayed in Figure~\ref{fig:zKmag_Ia}.\footnote{CC SNe are considered in Appendix~\ref{sec:cc_rcf}.} The joint distribution for detecting a host galaxy given its redshift and $M_{K_s}$, $\mathrm{RCF}(z, M_{K_s})$, along with the 1-dimensional probabilities, $\mathrm{RCF}(z, M_{K_s})$ and $\mathrm{RCF}(z, M_{K_s})$, are also shown in Figure~\ref{fig:zKmag_Ia}. Details for these calculations are presented in Appendix~\ref{sec:bayes}. Figure~\ref{fig:zKmag_Ia} confirms the intuitive results that the RCF is lower for higher redshift and intrinsically fainter galaxies.

In total, there are 42 host galaxies brighter than $M^\ast_{K_s}$,
and 2 of these did not have redshift entries in NED prior to SN
discovery. The first is the host of ASASSN-15ed,
MCG\,+09-27-087/SDSS\,J164825.26+505935.5, a bright, large
(40$^{\prime\prime}$) sprial galaxy. There is no SDSS
spectroscopic redshift but the SDSS photoz
estimate is $0.031\pm0.01$. The host galaxy has a $10.2\pm
1.1$\,mJy counterpart in NVSS. The second is the host of
ASASSN-15ub, CGCG\,314-006. Nominally there is no host redshift in
NED. However, a direct inspection of SDSS shows that a pair of
strongly interacting galaxies with a spectroscopic redshift of
0.032 (photoz of $0.027\pm 0.0076$) which can be compared to
$z_\mathrm{SN}=0.032$.

We restrict the analysis of sub-luminous galaxies to SNe Ia. In
the $K_s$ band, there are 163 host galaxies with $L < L_*$. Of
these 110 are listed in NED. Thus, as traced by SNe Ia, the
$\mathrm{RCF}=$ [67\%--73\%] (5\%--95\% confidence range).

\subsection{NUV/$u^\prime$ band}
\label{sec:NUVband}

As was the case for $K_s$ band, the median !NED$_z$ host galaxy is
$\sim$2\,mag fainter than the median NED$_z$ host galaxy for the
``UV'' band (Figure~\ref{fig:zUVmag_Ia}). The redshift distribution of NED$_z$ and !NED$_z$
galaxies in nearly identical in $K_s$ and ``UV'', with higher $z$
hosts more likely to not be included in NED. The peak of the !NED$_z$
sample ``UV'' luminosity is not as biased towards faint galaxies
as it is in the $K_s$ band. This conclusion bodes well for the PTF
Census of the Local Universe (CLU) H$\alpha$ survey which is
searching for nearby star-forming galaxies \citep{ckv+17a}.

In the ``UV'' band, 46 galaxies are brighter than $L_*$, and 5 of
those are !NED$_z$ galaxies. The host of ASASSN-15ed, MCG\,+09-27-087;
SDSS\,J164825.26+505935.5, is discussed above. The ASASSN-15ln
host, GALEXASC J225332.83+194232.9, is a 10$^{\prime\prime}$
spiral galaxy with no SDSS spectroscopic redshift but an SDSS
photoz of $0.022\pm 0.0092$. The host of ASASSN-15ho, 2MASXi
J0909234-044327, lies outside the SDSS footprint. The hosts of
ASASSN-15sh and ASASSN-15um, 2MASX\,J19320827-6226340 and
2MASX\,J05395948-8022191, respectively, lie in poorly studied
regions of the sky.

Again, we restrict the analysis of missing sub-luminous
galaxies to SNe Ia hosts. There are 140 galaxies with ``UV''
luminosity less than the corresponding $L_*$ value. Of these, 98
have redshift entries in NED. Thus, as traced by SNe Ia, the
$\mathrm{RCF}=$ [63\%--76\%].

\section{Increasing the Precision \& Accuracy of RCF} 
\label{sec:Census}

The ASAS-SN bright SN sample is attractive for measuring the RCF due to its host-unbiased approach and the essentially complete spectroscopic classification of all candidates that results from its shallow magnitude limit.  However, owing to the small sample size, the RCF estimates are limited by binomial errors. We undertook a similar analysis for a larger sample:
SNe candidates reported at the Transient Name Server (TNS)\footnote{\url{https://wis-tns.weizmann.ac.il/}} portal, during the period between January 2016 and June 2017.  
The sample spans a larger peak magnitude range relative to ASAS-SN, extending as faint as 20 mag. Nonetheless, it appears that follow up was obtained for most of the reported candidate SNe. The resulting sample size is 529 nearby 
($z\leq 0.05$) SNe.
We find that the RCF for this sample is similar to that derived for the ASAS-SN sample (Cassese \& Kulkarni, in prep). 

The field of optical time-domain astronomy is in a boom period, and much larger samples of nearby SNe can be expected given the Asteroid Terrestrial-impact Last Alert (ATLAS;
\citealt{t11}), PanSTARRS-1 \citep{wcl+16}, 
the Zwicky Transient Facility (ZTF; \citealt{dsb+16,bk17}), and 
upgraded ASAS-SN surveys. The limiting V-band magnitudes for these surveys range
from 17 to 21\,mag. Below, we consider the gains resulting from large SN samples.

To make this discussion concrete we consider a specific example, the ``Celestial Cinematography"  survey of ZTF \citep{bk17}. This survey aims to systematically cover 
a large fraction of the night sky ($\gtrsim 12,000\,\deg^2$) every three nights, in the $g$ and $R$
bands. The median 5-$\sigma$ detection limit, for a fixed 30\,s
exposure time, is 20.5\,mag.
The annual volumetric rate of $z \approx 0$ Type Ia and CC SNe is
$\mathcal{R}_{\rm Ia} \approx 3\times 10^4\,{\rm
Gpc^{-3}\,yr^{-1}}$ and $\mathcal{R}_{\rm CC} \approx 7 \times
10^4\,{\rm Gpc^{-3}\,yr^{-1}}$, respectively \citep{llc+11}. Based on a simulator built for ZTF, the expected
annual yield  for the above survey is [230, 460, 892, 1750] for peak magnitude of  [17.5, 18, 18.5,
19]\,mag, respectively (U.\ Feindt, pers.
comm.). 

Going forward we will assume a ``Bright Transient Survey'' (BTS) whose goal is to classify all extra-galactic transients whose peak magnitude is brighter than 18.5\,mag. A one year survey would result in a sample of nearly one thousand SNe Ia. With this sample, a regional RCF can be evaluated (e.g., high and intermediate Galactic latitude regions).
Next, the large and unbiased sample would allow for a number of other applications, including self-consistent checks of the dependence of the SN Ia rate on host type, which is frequently formulated as $\mathcal{P}\propto a\mathcal{M}+b\mathcal{S}$ \citep{sb05};
here, $a$ and $b$ are constants. Deviations will give us insight into a better formulation of the relationship between $\mathcal{P}$ and $\mathcal{M}$ and $\mathcal{S}$.

Such a large survey would, in its own right be interesting. For example, determining volumetric SN rates from untargeted, wide-field surveys requires the identification of all galaxies within a specified distance. The BTS measurement of the RCF will provide the correction factors needed to account for missing galaxies when calculating the volumetric rates. For example, the relative rate of SN\,2002cx-like (SNe Iax) to normal SNe Ia is wildly uncertain ($\approx 5-30$\%; e.g., \citealt{llc+11,fcc+13,mkc+17}), and the large sample from the BTS will substantially improve these estimates. Furthermore, such a large, low-redshift sample would be very valuable for Ia SN cosmography (e.g., \citealt{gaa+01,sjr+17}).

%CC

Next we address CC SNe. As noted in \S\ref{sec:Assessing} CC SNe exhibit a wide range in peak magnitude: $M_r$ ranging from  $-12$ to about
$-18$ mag \citep{tcd+14}. The BTS is well suited to determining the demographics of CC SNe. A survey complete to a flux-limit $V
\approx 18\,\mathrm{mag}$ will detect SNe peaking at $M_V = -12,
-15, \; \mathrm{and} \; -18$\,mag to a radius of $\sim$10, 40, and
160\,Mpc, respectively. The total number of CC SN detections will
sharply depend on the luminosity function. For instance,
\citet{tcd+14} suggest that the fraction of CC SNe fainter than
$-15\,\mathrm{mag}$ at peak is at least 24\% but can be as high as
$50\%$. In any case, BTS will allow us to
measure the luminosity function of CC events which is essential to
determine the volumetric rate of CC SNe. In turn, the latter is a
key element in our understanding of stars and the interstellar
medium \citep{hbk+11}. Finally, while CC SNe certainly track $\mathcal{S}$, it may be the case
that ``lesser" parameters, such as metallicity, change the mix of
CC SNe subtypes \citep{agk+2010,gsm+16,gbm17}. Again large-sample SN surveys may well have
sufficient diagnostic power to ferret out such connections.

It is increasingly evident that 
the primary limitation to SN surveys is limited by our ability to spectrally classify the SN candidates.   This load can be made bearable
by the use of two spectrographs: an ultra-low resolution
spectrometer tuned to classification and a standard low-resolution
spectrometer to get the redshift and gross spectrum of the host
galaxies. For the latter we note that within a few years not
merely highly- but \textit{supremely-}multiplexed spectrographs
(e.g.\ DESI\footnote{http://desi.lbl.gov},
PFS\footnote{http://pfs.ipmu.jp} and the planned AS4 project) will
be commissioned. These facilities, at very little cost (small
fractional allocation of fibers), can measure the redshifts of
host galaxies of SNe on an industrial scale. The same
highly-multiplexed spectrographs will likely be pressed into
surveys more ambitious than SDSS or 6dF, leading to more complete
catalogs of galaxies in the nearby Universe.

\acknowledgements   
We thank A.\ Goobar, U.\ Feindt, C.\ Pankow, M.\ Kasliwal, P.\ Nugent, E.\ O.\ Ofek,  E.\ S.\ Phinney, K.\ Taggart and H.\ Vedantham for inputs and helpful discussions.

AAM is funded by the Large Synoptic Survey Telescope Corporation in support of the Data Science Fellowship Program.

%\software{\texttt{emcee} \citep{fhl+13}, 
%		  \texttt{corner} \citep{corner},
%          \texttt{matplotlib} \citep{Hunter2007,matplotlib222}, 
%          \texttt{scipy} \citep{scipy},
%          \texttt{pandas} \citep{pandas}}

\appendix

\section{Conditional Probability of the RCF}\label{sec:bayes}

We aim to characterize the RCF as a function of redshift, $z$, and host galaxy luminosity, where we use either $M_{K_s, \mathrm{host}}$ or $M_{UV, \mathrm{host}}$ as a proxy. To do so, we model the data $X$ with the Bernoulli distribution
\begin{equation}
	X \sim \operatorname{Bern}(p),
\end{equation}
where $p$ is parameterized with a logistic function with dependence on both redshift $z$ and host galaxy luminosity:
\begin{equation}
p(z, M, \theta) = \frac{1}{1 + \exp(az + bM - c)},
\end{equation}
with host-galaxy absolute magnitude $M$, and $\theta$ representing the model parameters: $a$, $b$, and $c$, which need to be determined. The precise analytic dependence of $p$ on $z$ and $M$ may not be logistic, however, the purpose of this exercise is to provide a general sense for how the RCF relies on $z$ and $M$. The logistic function is ideal for this general purpose.

From here it follows that the probability of a host galaxy having a previously cataloged redshift is:
\begin{equation}
Pr(q) = \left\{
\begin{array}{ll}
      p(z, M, \theta), & \mathrm{if} \, q=\mathrm{NED}_z \\
      1-p(z, M, \theta), & \mathrm{if} \, q=\mathrm{!NED}_z
\end{array}\right.
\label{eqn:bern}
\end{equation}
and the likelihood of the observations given the data and model parameters is:
\begin{equation}
Pr({q_k}\,|\,{z_k}, M_K, \theta) = \prod_{k=1}^K p(z_k, M_k, \theta)^{q_k} \, (1-p(z_k, M_k, \theta))^{1 - q_k},
\label{eqn:likelihood}
\end{equation}
where $k$ represents the individual observations and $q_k = 1$ for $\mathrm{NED}_z$ galaxies and $q_k = 0$ for $\mathrm{!NED}_z$ galaxies.

From Bayes' theorem, we can multiply the likelihood by a prior, $Pr(\theta)$, and use Markov Chain Monte Carlo (MCMC) techniques to sample from the posterior $Pr(\theta\,|\,{q_k}, {z_k}, M_K)$ in order to constrain the model parameters $\theta$. We use the \texttt{emcee} package \citep{fhl+13} to implement our MCMC sampling of the posterior. For $a$ and $b$ we adopt flat priors bounded between 0 and $10^{6}$. For $c$ we adopt a flat prior between $-100$ and $100$. Following the MCMC sampling, we find that there is a strong covariance between $b$ and $c$, while $a$ is relatively independent of $b$ and $c$. The shading in Figures~\ref{fig:zKmag_Ia} and~\ref{fig:zUVmag_Ia} shows $p(z, M, \theta)$ for the maximum a posteriori sample from the MCMC sampling.

We additionally wish to constrain the behavior of the RCF as a function of either the host redshift, $z$, or host galaxy luminosity. We do this separately from the analysis above, while using the same MCMC procedure with $p$ in Equations~\ref{eqn:bern} and~\ref{eqn:likelihood} replaced by
\begin{equation}
p(z, \theta) = \frac{1}{1 + \exp(az - c)},
\label{eqn:p_z}
\end{equation}
for redshift, and 
\begin{equation}
p(M, \theta) = \frac{1}{1 + \exp(bM - c)},
\end{equation}
for host galaxy luminosity (where, again, we use absolute magnitude $M$ as a proxy). The results of this procedure are shown in the side panels of Figures~\ref{fig:zKmag_Ia} and~\ref{fig:zUVmag_Ia}. In these panels the solid lines show the median value of $p(z)$, $\mathrm{RCF}(z)$ in the Figures, and $p(M)$, $\mathrm{RCF}(M)$ in the Figures, from all the posterior samples, while the shaded region shows the 90\% credible regions for $p(z)$ and $p(M)$ from the posterior samples. We close by noting that the current dataset provides weak constraints on $a$ in Equation~\ref{eqn:p_z}, but these constraints will be greatly improved by the BTS which will include a significantly larger sample and extend to higher redshifts.

\section{RCF as Traced by CC SNe}\label{sec:cc_rcf}

\begin{figure*}[hbp] 
  \centering
  \includegraphics[width=6in]{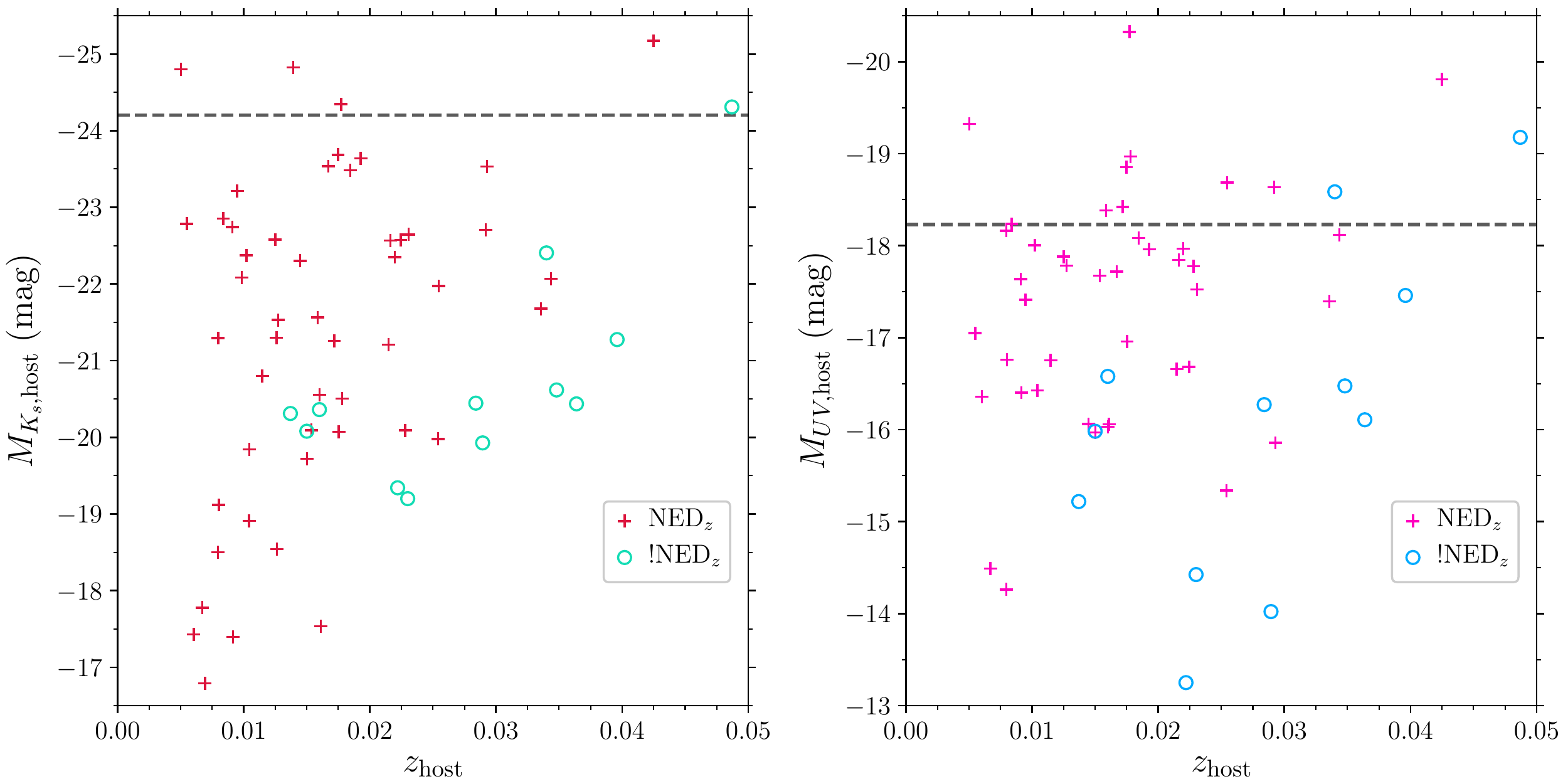} 
  \caption{\textit{Left}: Absolute $K_s$-band magnitude, $M_{K_s,\mathrm{host}}$, vs.\ redshift, $z_\mathrm{host}$, for the host galaxies of CC SNe in \textit{A1} and \textit{A2}. Galaxies with redshift entries in NED are shown as red pluses, while those lacking redshifts (!NED$_z$) are shown as teal circles. The horizontal dashed line shows $M_{K_s}^\ast$. \textit{Right}: Absolute ``UV''-band magnitude, $M_{UV,\mathrm{host}}$, vs.\ redshift, $z_\mathrm{host}$, for the host galaxies of CC SNe in \textit{A1} and \textit{A2}. Galaxies with redshift entries in NED are shown as magenta pluses, while those lacking redshifts (!NED$_z$) are shown as blue circles. The horizontal dashed line shows $M_{UV}^\ast$.
  }
\label{fig:ccsne}
\end{figure*}

The CC SNe samples in A1 and A2 are insufficient to meaningfully constrain the RCF as a function of redshift or host galaxy absolute magnitude. Furthermore, CC SNe only trace star formation, meaning they do not probe passive galaxies, so we have excluded them from the analysis in the main text. Nevertheless, for completeness, we show the host galaxies for CC SNe in Figure~\ref{fig:ccsne}.

\begin{thebibliography}{}
\expandafter\ifx\csname natexlab\endcsname\relax\def\natexlab#1{#1}\fi

\bibitem[{Abbott {et~al.}(2017a)Abbott, Abbott, Abbott, Acernese, Ackley,
  Adams, Adams, Addesso, Adhikari, Adya, Affeldt, Afrough, Agarwal, Agathos,
  Agatsuma, Aggarwal, Aguiar, Aiello, Ain, Ajith, Allen, Allen, Allocca, Altin,
  Amato, Ananyeva, Anderson, Anderson, Angelova, Antier, Appert, Arai, Araya,
  Areeda, Arnaud, Arun, Ascenzi, Ashton, Ast, Aston, Astone, Atallah, Aufmuth,
  Aulbert, AultONeal, Austin, Avila-Alvarez, Babak, Bacon, Bader, Bae, Bailes,
  Baker, Baldaccini, Ballardin, Ballmer, Banagiri, Barayoga, Barclay, Barish,
  Barker, Barkett, Barone, Barr, Barsotti, Barsuglia, Barta, Barthelmy,
  Bartlett, Bartos, Bassiri, Basti, Batch, Bawaj, Bayley, Bazzan, B\'ecsy,
  Beer, Bejger, Belahcene, Bell, Berger, Bergmann, Bernuzzi, Bero, Berry,
  Bersanetti, Bertolini, Betzwieser, Bhagwat, Bhandare, Bilenko, Billingsley,
  Billman, Birch, Birney, Birnholtz, Biscans, Biscoveanu, Bisht, Bitossi,
  Biwer, Bizouard, Blackburn, Blackman, Blair, Blair, Blair, Bloemen, Bock,
  Bode, Boer, Bogaert, Bohe, Bondu, Bonilla, Bonnand, Boom, Bork, Boschi, Bose,
  Bossie, Bouffanais, Bozzi, Bradaschia, Brady, Branchesi, Brau, Briant,
  Brillet, Brinkmann, Brisson, Brockill, Broida, Brooks, Brown, Brown, Brunett,
  Buchanan, Buikema, Bulik, Bulten, Buonanno, Buskulic, Buy, Byer, Cabero,
  Cadonati, Cagnoli, Cahillane, Calder\'on~Bustillo, Callister, Calloni, Camp,
  Canepa, Canizares, Cannon, Cao, Cao, Capano, Capocasa, Carbognani, Caride,
  Carney, Carullo, Casanueva~Diaz, Casentini, Caudill, Cavagli\`a, Cavalier,
  Cavalieri, Cella, Cepeda, Cerd\'a-Dur\'an, Cerretani, Cesarini, Chamberlin,
  Chan, Chao, Charlton, Chase, Chassande-Mottin, Chatterjee, Chatziioannou,
  Cheeseboro, Chen, Chen, Chen, Cheng, Chia, Chincarini, Chiummo, Chmiel, Cho,
  Cho, Chow, Christensen, Chu, Chua, Chua, Chung, Chung, Ciani, Ciolfi,
  Cirelli, Cirone, Clara, Clark, Clearwater, Cleva, Cocchieri, Coccia, Cohadon,
  Cohen, Colla, Collette, Cominsky, Constancio, Conti, Cooper, Corban, Corbitt,
  Cordero-Carri\'on, Corley, Cornish, Corsi, Cortese, Costa, Coughlin,
  Coughlin, Coulon, Countryman, Couvares, Covas, Cowan, Coward, Cowart, Coyne,
  Coyne, Creighton, Creighton, Cripe, Crowder, Cullen, Cumming, Cunningham,
  Cuoco, Dal~Canton, D\'alya, Danilishin, D'Antonio, Danzmann, Dasgupta,
  Da~Silva~Costa, Dattilo, Dave, Davier, Davis, Daw, Day, De, DeBra, Degallaix,
  De~Laurentis, Del\'eglise, Del~Pozzo, Demos, Denker, Dent, De~Pietri,
  Dergachev, De~Rosa, DeRosa, De~Rossi, DeSalvo, de~Varona, Devenson,
  Dhurandhar, D\'{\i}az, Dietrich, Di~Fiore, Di~Giovanni, Di~Girolamo,
  Di~Lieto, Di~Pace, Di~Palma, Di~Renzo, Doctor, Dolique, Donovan, Dooley,
  Doravari, Dorrington, Douglas, Dovale~\'Alvarez, Downes, Drago,
  Dreissigacker, Driggers, Du, Ducrot, Dudi, Dupej, Dwyer, Edo, Edwards,
  Effler, Eggenstein, Ehrens, Eichholz, Eikenberry, Eisenstein, Essick,
  Estevez, Etienne, Etzel, Evans, Evans, Factourovich, Fafone, Fair, Fairhurst,
  Fan, Farinon, Farr, Farr, Fauchon-Jones, Favata, Fays, Fee, Fehrmann, Feicht,
  Fejer, Fernandez-Galiana, Ferrante, Ferreira, Ferrini, Fidecaro, Finstad,
  Fiori, Fiorucci, Fishbach, Fisher, Fitz-Axen, Flaminio, Fletcher, Fong, Font,
  Forsyth, Forsyth, Fournier, Frasca, Frasconi, Frei, Freise, Frey, Frey,
  Fries, Fritschel, Frolov, Fulda, Fyffe, Gabbard, Gadre, Gaebel, Gair,
  Gammaitoni, Ganija, Gaonkar, Garcia-Quiros, Garufi, Gateley, Gaudio, Gaur,
  Gayathri, Gehrels, Gemme, Genin, Gennai, George, George, Gergely, Germain,
  Ghonge, Ghosh, Ghosh, Ghosh, Giaime, Giardina, Giazotto, Gill, Glover, Goetz,
  Goetz, Gomes, Goncharov, Gonz\'alez, Gonzalez~Castro, Gopakumar, Gorodetsky,
  Gossan, Gosselin, Gouaty, Grado, Graef, Granata, Grant, Gras, Gray, Greco,
  Green, Gretarsson, Groot, Grote, Grunewald, Gruning, Guidi, Guo, Gupta,
  Gupta, Gushwa, Gustafson, Gustafson, Halim, Hall, Hall, Hamilton, Hammond,
  Haney, Hanke, Hanks, Hanna, Hannam, Hannuksela, Hanson, Hardwick, Harms,
  Harry, Harry, Hart, Haster, Haughian, Healy, Heidmann, Heintze, Heitmann,
  Hello, Hemming, Hendry, Heng, Hennig, Heptonstall, Heurs, Hild, Hinderer, Ho,
  Hoak, Hofman, Holt, Holz, Hopkins, Horst, Hough, Houston, Howell, Hreibi, Hu,
  Huerta, Huet, Hughey, Husa, Huttner, Huynh-Dinh, Indik, Inta, Intini, Isa,
  Isac, Isi, Iyer, Izumi, Jacqmin, Jani, Jaranowski, Jawahar,
  Jim\'enez-Forteza, Johnson, Johnson-McDaniel, Jones, Jones, Jonker, Ju,
  Junker, Kalaghatgi, Kalogera, Kamai, Kandhasamy, Kang, Kanner, Kapadia,
  Karki, Karvinen, Kasprzack, Kastaun, Katolik, Katsavounidis, Katzman, Kaufer,
  Kawabe, K\'ef\'elian, Keitel, Kemball, Kennedy, Kent, Key, Khalili, Khan,
  Khan, Khan, Khazanov, Kijbunchoo, Kim, Kim, Kim, Kim, Kim, Kim, Kimbrell,
  King, King, Kinley-Hanlon, Kirchhoff, Kissel, Kleybolte, Klimenko, Knowles,
  Koch, Koehlenbeck, Koley, Kondrashov, Kontos, Korobko, Korth, Kowalska,
  Kozak, Kr\"amer, Kringel, Krishnan, Kr\'olak, Kuehn, Kumar, Kumar, Kumar,
  Kuo, Kutynia, Kwang, Lackey, Lai, Landry, Lang, Lange, Lantz, Lanza, Larson,
  Lartaux-Vollard, Lasky, Laxen, Lazzarini, Lazzaro, Leaci, Leavey, Lee, Lee,
  Lee, Lee, Lee, Lehmann, Lenon, Leon, Leonardi, Leroy, Letendre, Levin, Li,
  Linker, Littenberg, Liu, Liu, Lo, Lockerbie, London, Lord, Lorenzini,
  Loriette, Lormand, Losurdo, Lough, Lousto, Lovelace, L\"uck, Lumaca,
  Lundgren, Lynch, Ma, Macas, Macfoy, Machenschalk, MacInnis, Macleod, Maga\~na
  Hernandez, Maga\~na Sandoval, Maga\~na Zertuche, Magee, Majorana, Maksimovic,
  Man, Mandic, Mangano, Mansell, Manske, Mantovani, Marchesoni, Marion,
  M\'arka, M\'arka, Markakis, Markosyan, Markowitz, Maros, Marquina, Marsh,
  Martelli, Martellini, Martin, Martin, Martynov, Marx, Mason, Massera,
  Masserot, Massinger, Masso-Reid, Mastrogiovanni, Matas, Matichard, Matone,
  Mavalvala, Mazumder, McCarthy, McClelland, McCormick, McCuller, McGuire,
  McIntyre, McIver, McManus, McNeill, McRae, McWilliams, Meacher, Meadors,
  Mehmet, Meidam, Mejuto-Villa, Melatos, Mendell, Mercer, Merilh, Merzougui,
  Meshkov, Messenger, Messick, Metzdorff, Meyers, Miao, Michel, Middleton,
  Mikhailov, Milano, Miller, Miller, Miller, Millhouse, Milovich-Goff,
  Minazzoli, Minenkov, Ming, Mishra, Mitra, Mitrofanov, Mitselmakher,
  Mittleman, Moffa, Moggi, Mogushi, Mohan, Mohapatra, Molina, Montani, Moore,
  Moraru, Moreno, Morisaki, Morriss, Mours, Mow-Lowry, Mueller, Muir,
  Mukherjee, Mukherjee, Mukherjee, Mukund, Mullavey, Munch, Mu\~niz, Muratore,
  Murray, Nagar, Napier, Nardecchia, Naticchioni, Nayak, Neilson, Nelemans,
  Nelson, Nery, Neunzert, Nevin, Newport, Newton, Ng, Nguyen, Nguyen, Nichols,
  Nielsen, Nissanke, Nitz, Noack, Nocera, Nolting, North, Nuttall, Oberling,
  O'Dea, Ogin, Oh, Oh, Ohme, Okada, Oliver, Oppermann, Oram, O'Reilly,
  Ormiston, Ortega, O'Shaughnessy, Ossokine, Ottaway, Overmier, Owen, Pace,
  Page, Page, Pai, Pai, Palamos, Palashov, Palomba, Pal-Singh, Pan, Pan, Pang,
  Pang, Pankow, Pannarale, Pant, Paoletti, Paoli, Papa, Parida, Parker,
  Pascucci, Pasqualetti, Passaquieti, Passuello, Patil, Patricelli, Pearlstone,
  Pedraza, Pedurand, Pekowsky, Pele, Penn, Perez, Perreca, Perri, Pfeiffer,
  Phelps, Piccinni, Pichot, Piergiovanni, Pierro, Pillant, Pinard, Pinto,
  Pirello, Pitkin, Poe, Poggiani, Popolizio, Porter, Post, Powell, Prasad,
  Pratt, Pratten, Predoi, Prestegard, Prijatelj, Principe, Privitera, Prix,
  Prodi, Prokhorov, Puncken, Punturo, Puppo, P\"urrer, Qi, Quetschke, Quintero,
  Quitzow-James, Raab, Rabeling, Radkins, Raffai, Raja, Rajan, Rajbhandari,
  Rakhmanov, Ramirez, Ramos-Buades, Rapagnani, Raymond, Razzano, Read,
  Regimbau, Rei, Reid, Reitze, Ren, Reyes, Ricci, Ricker, Rieger, Riles, Rizzo,
  Robertson, Robie, Robinet, Rocchi, Rolland, Rollins, Roma, Romano, Romano,
  Romel, Romie, Rosi\ifmmode~\acute{n}\else \'{n}\fi{}ska, Ross, Rowan,
  R\"udiger, Ruggi, Rutins, Ryan, Sachdev, Sadecki, Sadeghian, Sakellariadou,
  Salconi, Saleem, Salemi, Samajdar, Sammut, Sampson, Sanchez, Sanchez,
  Sanchis-Gual, Sandberg, Sanders, Sassolas, Sathyaprakash, Saulson, Sauter,
  Savage, Sawadsky, Schale, Scheel, Scheuer, Schmidt, Schmidt, Schnabel,
  Schofield, Sch\"onbeck, Schreiber, Schuette, Schulte, Schutz, Schwalbe,
  Scott, Scott, Seidel, Sellers, Sengupta, Sentenac, Sequino, Sergeev,
  Shaddock, Shaffer, Shah, Shahriar, Shaner, Shao, Shapiro, Shawhan, Sheperd,
  Shoemaker, Shoemaker, Siellez, Siemens, Sieniawska, Sigg, Silva, Singer,
  Singh, Singhal, Sintes, Slagmolen, Smith, Smith, Smith, Somala, Son,
  Sonnenberg, Sorazu, Sorrentino, Souradeep, Spencer, Srivastava, Staats,
  Staley, Steinke, Steinlechner, Steinlechner, Steinmeyer, Stevenson, Stone,
  Stops, Strain, Stratta, Strigin, Strunk, Sturani, Stuver, Summerscales, Sun,
  Sunil, Suresh, Sutton, Swinkels, Szczepa\ifmmode~\acute{n}\else
  \'{n}\fi{}czyk, Tacca, Tait, Talbot, Talukder, Tanner, T\'apai, Taracchini,
  Tasson, Taylor, Taylor, Tewari, Theeg, Thies, Thomas, Thomas, Thomas, Thorne,
  Thorne, Thrane, Tiwari, Tiwari, Tokmakov, Toland, Tonelli, Tornasi,
  Torres-Forn\'e, Torrie, T\"oyr\"a, Travasso, Traylor, Trinastic, Tringali,
  Trozzo, Tsang, Tse, Tso, Tsukada, Tsuna, Tuyenbayev, Ueno, Ugolini,
  Unnikrishnan, Urban, Usman, Vahlbruch, Vajente, Valdes, Vallisneri, van
  Bakel, van Beuzekom, van~den Brand, Van Den~Broeck, Vander-Hyde, van~der
  Schaaf, van Heijningen, van Veggel, Vardaro, Varma, Vass, Vas\'uth, Vecchio,
  Vedovato, Veitch, Veitch, Venkateswara, Venugopalan, Verkindt, Vetrano,
  Vicer\'e, Viets, Vinciguerra, Vine, Vinet, Vitale, Vo, Vocca, Vorvick,
  Vyatchanin, Wade, Wade, Wade, Walet, Walker, Wallace, Walsh, Wang, Wang,
  Wang, Wang, Wang, Ward, Warner, Was, Watchi, Weaver, Wei, Weinert, Weinstein,
  Weiss, Wen, Wessel, We\ss{}els, Westerweck, Westphal, Wette, Whelan,
  Whitcomb, Whiting, Whittle, Wilken, Williams, Williams, Williamson, Willis,
  Willke, Wimmer, Winkler, Wipf, Wittel, Woan, Woehler, Wofford, Wong, Worden,
  Wright, Wu, Wysocki, Xiao, Yamamoto, Yancey, Yang, Yap, Yazback, Yu, Yu,
  Yvert, Zadro\ifmmode~\dot{z}\else \.{z}\fi{}ny, Zanolin, Zelenova, Zendri,
  Zevin, Zhang, Zhang, Zhang, Zhang, Zhao, Zhou, Zhou, Zhu, Zhu, Zimmerman,
  Zucker, \& Zweizig}]{aaa+17}
Abbott, B.~P., Abbott, R., Abbott, T.~D., {et~al.} 2017a, Phys. Rev. Lett.,
  119, 161101

\bibitem[{{Abbott} {et~al.}(2017b){Abbott}, {Abbott}, {Abbott}, {Acernese},
  {Ackley}, {Adams}, {Adams}, {Addesso}, {Adhikari}, {Adya}, \&
  et~al.}]{aaa+17a}
{Abbott}, B.~P., {Abbott}, R., {Abbott}, T.~D., {et~al.} 2017b, \apjl, 848, L12

\bibitem[{{Arcavi} {et~al.}(2010){Arcavi}, {Gal-Yam}, {Kasliwal}, {Quimby},
  {Ofek}, {Kulkarni}, {Nugent}, {Cenko}, {Bloom}, {Sullivan}, {Howell},
  {Poznanski}, {Filippenko}, {Law}, {Hook}, {J{\"o}nsson}, {Blake}, {Cooke},
  {Dekany}, {Rahmer}, {Hale}, {Smith}, {Zolkower}, {Velur}, {Walters},
  {Henning}, {Bui}, {McKenna}, \& {Jacobsen}}]{agk+2010}
{Arcavi}, I., {Gal-Yam}, A., {Kasliwal}, M.~M., {et~al.} 2010, \apj, 721, 777

\bibitem[{{Bellm} \& {Kulkarni}(2017)}]{bk17}
{Bellm}, E., \& {Kulkarni}, S. 2017, Nature Astronomy, 1, 0071

\bibitem[{{Blanton} \& {Moustakas}(2009)}]{bm09}
{Blanton}, M.~R., \& {Moustakas}, J. 2009, \araa, 47, 159

\bibitem[{{Cole} {et~al.}(2001){Cole}, {Norberg}, {Baugh}, {Frenk},
  {Bland-Hawthorn}, {Bridges}, {Cannon}, {Colless}, {Collins}, {Couch},
  {Cross}, {Dalton}, {De Propris}, {Driver}, {Efstathiou}, {Ellis},
  {Glazebrook}, {Jackson}, {Lahav}, {Lewis}, {Lumsden}, {Maddox}, {Madgwick},
  {Peacock}, {Peterson}, {Sutherland}, \& {Taylor}}]{cnb+01}
{Cole}, S., {Norberg}, P., {Baugh}, C.~M., {et~al.} 2001, \mnras, 326, 255

\bibitem[{{Cook} {et~al.}(2017){Cook}, {Kasliwal}, {Van Sistine}, {Kaplan},
  {Sutter}, {Kupfer}, {Shupe}, {Laher}, {Masci}, {Dale}, {Sesar}, {Brady},
  {Yan}, \& {Ofek}}]{ckv+17a}
{Cook}, D.~O., {Kasliwal}, M.~M., {Van Sistine}, A., {et~al.} 2017, ArXiv
  e-prints, arXiv:1710.05016

\bibitem[{{Dekany} {et~al.}(2016){Dekany}, {Smith}, {Belicki}, {Delacroix},
  {Duggan}, {Feeney}, {Hale}, {Kaye}, {Milburn}, {Murphy}, {Porter}, {Reiley},
  {Riddle}, {Rodriguez}, \& {Bellm}}]{dsb+16}
{Dekany}, R., {Smith}, R.~M., {Belicki}, J., {et~al.} 2016, in \procspie, Vol.
  9908, Ground-based and Airborne Instrumentation for Astronomy VI, 99085M

\bibitem[{Droettboom {et~al.}(2018)Droettboom, Caswell, Hunter, Firing,
  Nielsen, Lee, Andrade, Varoquaux, Stansby, Root, Elson, Dale, {Jae-Joon Lee},
  May, Sepp\"{a}nen, Klymak, McDougall, Straw, Hobson, {Cgohlke}, Yu, Ma,
  Vincent, Silvester, Moad, Katins, Kniazev, Hoffmann, Ariza, \&
  W\"{u}rtz}]{matplotlib222}
Droettboom, M., Caswell, T.~A., Hunter, J., {et~al.} 2018,
  Matplotlib/Matplotlib V2.2.2,  Zenodo, doi:10.5281/zenodo.1202077

\bibitem[{{Foley} {et~al.}(2013){Foley}, {Challis}, {Chornock},
  {Ganeshalingam}, {Li}, {Marion}, {Morrell}, {Pignata}, {Stritzinger},
  {Silverman}, {Wang}, {Anderson}, {Filippenko}, {Freedman}, {Hamuy}, {Jha},
  {Kirshner}, {McCully}, {Persson}, {Phillips}, {Reichart}, \&
  {Soderberg}}]{fcc+13}
{Foley}, R.~J., {Challis}, P.~J., {Chornock}, R., {et~al.} 2013, \apj, 767, 57

\bibitem[{Foreman-Mackey(2016)}]{corner}
Foreman-Mackey, D. 2016, The Journal of Open Source Software, 24,
  doi:10.21105/joss.00024

\bibitem[{{Foreman-Mackey} {et~al.}(2013){Foreman-Mackey}, {Hogg}, {Lang}, \&
  {Goodman}}]{fhl+13}
{Foreman-Mackey}, D., {Hogg}, D.~W., {Lang}, D., \& {Goodman}, J. 2013, \pasp,
  125, 306

\bibitem[{{Galbany} {et~al.}(2016){Galbany}, {Stanishev}, {Mour{\~a}o},
  {Rodrigues}, {Flores}, {Walcher}, {S{\'a}nchez}, {Garc{\'{\i}}a-Benito},
  {Mast}, {Badenes}, {Gonz{\'a}lez Delgado}, {Kehrig}, {Lyubenova}, {Marino},
  {Moll{\'a}}, {Meidt}, {P{\'e}rez}, {van de Ven}, \&
  {V{\'{\i}}lchez}}]{gsm+16}
{Galbany}, L., {Stanishev}, V., {Mour{\~a}o}, A.~M., {et~al.} 2016, \aap, 591,
  A48

\bibitem[{{Gehrels} {et~al.}(2016){Gehrels}, {Cannizzo}, {Kanner}, {Kasliwal},
  {Nissanke}, \& {Singer}}]{gck+16}
{Gehrels}, N., {Cannizzo}, J.~K., {Kanner}, J., {et~al.} 2016, \apj, 820, 136

\bibitem[{{Goliath} {et~al.}(2001){Goliath}, {Amanullah}, {Astier}, {Goobar},
  \& {Pain}}]{gaa+01}
{Goliath}, M., {Amanullah}, R., {Astier}, P., {Goobar}, A., \& {Pain}, R. 2001,
  \aap, 380, 6

\bibitem[{{Graur} {et~al.}(2017){Graur}, {Bianco}, {Modjaz}, {Shivvers},
  {Filippenko}, {Li}, \& {Smith}}]{gbm17}
{Graur}, O., {Bianco}, F.~B., {Modjaz}, M., {et~al.} 2017, \apj, 837, 121

\bibitem[{{Holoien} {et~al.}(2017{\natexlab{a}}){Holoien}, {Stanek},
  {Kochanek}, {Shappee}, {Prieto}, {Brimacombe}, {Bersier}, {Bishop}, {Dong},
  {Brown}, {Danilet}, {Simonian}, {Basu}, {Beacom}, {Falco}, {Pojmanski},
  {Skowron}, {Wo{\'z}niak}, {{\'A}vila}, {Conseil}, {Contreras}, {Cruz},
  {Fern{\'a}ndez}, {Koff}, {Guo}, {Herczeg}, {Hissong}, {Hsiao}, {Jose},
  {Kiyota}, {Long}, {Monard}, {Nicholls}, {Nicolas}, \& {Wiethoff}}]{hsk+17}
{Holoien}, T.~W.-S., {Stanek}, K.~Z., {Kochanek}, C.~S., {et~al.}
  2017{\natexlab{a}}, \mnras, 464, 2672

\bibitem[{{Holoien} {et~al.}(2017{\natexlab{b}}){Holoien}, {Brown}, {Stanek},
  {Kochanek}, {Shappee}, {Prieto}, {Dong}, {Brimacombe}, {Bishop}, {Basu},
  {Beacom}, {Bersier}, {Chen}, {Danilet}, {Falco}, {Godoy-Rivera}, {Goss},
  {Pojmanski}, {Simonian}, {Skowron}, {Thompson}, {Wo{\'z}niak}, {{\'A}vila},
  {Bock}, {Carballo}, {Conseil}, {Contreras}, {Cruz}, {And{\'u}jar}, {Guo},
  {Hsiao}, {Kiyota}, {Koff}, {Krannich}, {Madore}, {Marples}, {Masi},
  {Morrell}, {Monard}, {Munoz-Mateos}, {Nicholls}, {Nicolas}, {Wagner}, \&
  {Wiethoff}}]{hbs+17}
{Holoien}, T.~W.-S., {Brown}, J.~S., {Stanek}, K.~Z., {et~al.}
  2017{\natexlab{b}}, \mnras, 467, 1098

\bibitem[{{Horiuchi} {et~al.}(2011){Horiuchi}, {Beacom}, {Kochanek}, {Prieto},
  {Stanek}, \& {Thompson}}]{hbk+11}
{Horiuchi}, S., {Beacom}, J.~F., {Kochanek}, C.~S., {et~al.} 2011, \apj, 738,
  154

\bibitem[{Hunter(2007)}]{Hunter2007}
Hunter, J.~D. 2007, Computing In Science \& Engineering, 9, 90

\bibitem[{{Jha} {et~al.}(2007){Jha}, {Riess}, \& {Kirshner}}]{jrk07}
{Jha}, S., {Riess}, A.~G., \& {Kirshner}, R.~P. 2007, \apj, 659, 122

\bibitem[{Jones {et~al.}(2001--)Jones, Oliphant, Peterson, {et~al.}}]{scipy}
Jones, E., Oliphant, T., Peterson, P., {et~al.} 2001--, {SciPy}: Open source
  scientific tools for {Python}, , , [Online; accessed <today>]

\bibitem[{{Karachentsev} {et~al.}(2013){Karachentsev}, {Makarov}, \&
  {Kaisina}}]{kmk13}
{Karachentsev}, I.~D., {Makarov}, D.~I., \& {Kaisina}, E.~I. 2013, \aj, 145,
  101

\bibitem[{{Kasliwal} {et~al.}(2016){Kasliwal}, {Cenko}, {Singer}, {Corsi},
  {Cao}, {Barlow}, {Bhalerao}, {Bellm}, {Cook}, {Duggan}, {Ferretti}, {Frail},
  {Horesh}, {Kendrick}, {Kulkarni}, {Lunnan}, {Palliyaguru}, {Laher}, {Masci},
  {Manulis}, {Miller}, {Nugent}, {Perley}, {Prince}, {Quimby}, {Rana},
  {Rebbapragada}, {Sesar}, {Singhal}, {Surace}, \& {Van Sistine}}]{ksc+16}
{Kasliwal}, M.~M., {Cenko}, S.~B., {Singer}, L.~P., {et~al.} 2016, \apjl, 824,
  L24

\bibitem[{{Li} {et~al.}(2011){Li}, {Leaman}, {Chornock}, {Filippenko},
  {Poznanski}, {Ganeshalingam}, {Wang}, {Modjaz}, {Jha}, {Foley}, \&
  {Smith}}]{llc+11}
{Li}, W., {Leaman}, J., {Chornock}, R., {et~al.} 2011, \mnras, 412, 1441

\bibitem[{{Li} {et~al.}(2000){Li}, {Filippenko}, {Treffers}, {Friedman},
  {Halderson}, {Johnson}, {King}, {Modjaz}, {Papenkova}, {Sato}, \&
  {Shefler}}]{lft+00}
{Li}, W.~D., {Filippenko}, A.~V., {Treffers}, R.~R., {et~al.} 2000, in American
  Institute of Physics Conference Series, Vol. 522, American Institute of
  Physics Conference Series, ed. S.~S. {Holt} \& W.~W. {Zhang}, 103--106

\bibitem[{{McCray}(1993)}]{mccray93}
{McCray}, R. 1993, \araa, 31, 175

\bibitem[{McKinney(2010)}]{pandas}
McKinney, W. 2010, in Proceedings of the 9th Python in Science Conference, ed.
  S.~van~der Walt \& J.~Millman, 51 -- 56

\bibitem[{{Miller} {et~al.}(2017){Miller}, {Kasliwal}, {Cao}, {Adams},
  {Goobar}, {Kne{\v z}evi{\'c}}, {Laher}, {Lunnan}, {Masci}, {Nugent},
  {Perley}, {Petrushevska}, {Quimby}, {Rebbapragada}, {Sollerman}, {Taddia}, \&
  {Kulkarni}}]{mkc+17}
{Miller}, A.~A., {Kasliwal}, M.~M., {Cao}, Y., {et~al.} 2017, \apj, 848, 59

\bibitem[{{Paturel} {et~al.}(2003){Paturel}, {Petit}, {Prugniel}, {Theureau},
  {Rousseau}, {Brouty}, {Dubois}, \& {Cambr{\'e}sy}}]{ppt+03}
{Paturel}, G., {Petit}, C., {Prugniel}, P., {et~al.} 2003, \aap, 412, 45

\bibitem[{{Scannapieco} \& {Bildsten}(2005)}]{sb05}
{Scannapieco}, E., \& {Bildsten}, L. 2005, \apjl, 629, L85

\bibitem[{{Scolnic} {et~al.}(2017){Scolnic}, {Jones}, {Rest}, {Pan},
  {Chornock}, {Foley}, {Huber}, {Kessler}, {Narayan}, {Riess}, {Rodney},
  {Berger}, {Brout}, {Challis}, {Drout}, {Finkbeiner}, {Lunnan}, {Kirshner},
  {Sanders}, {Schlafly}, {Smartt}, {Stubbs}, {Tonry}, {Wood-Vasey}, {Foley},
  {Hand}, {Johnson}, {Burgett}, {Chambers}, {Draper}, {Hodapp}, {Kaiser},
  {Kudritzki}, {Magnier}, {Metcalfe}, {Bresolin}, {Gall}, {Kotak}, {McCrum}, \&
  {Smith}}]{sjr+17}
{Scolnic}, D.~M., {Jones}, D.~O., {Rest}, A., {et~al.} 2017, ArXiv e-prints,
  arXiv:1710.00845

\bibitem[{{Smartt} {et~al.}(2016){Smartt}, {Chambers}, {Smith}, {Huber},
  {Young}, {Cappellaro}, {Wright}, {Coughlin}, {Schultz}, {Denneau},
  {Flewelling}, {Heinze}, {Magnier}, {Primak}, {Rest}, {Sherstyuk}, {Stalder},
  {Stubbs}, {Tonry}, {Waters}, {Willman}, {Anderson}, {Baltay}, {Botticella},
  {Campbell}, {Dennefeld}, {Chen}, {Della Valle}, {Elias-Rosa}, {Fraser},
  {Inserra}, {Kankare}, {Kotak}, {Kupfer}, {Harmanen}, {Galbany}, {Gal-Yam},
  {Le Guillou}, {Lyman}, {Maguire}, {Mitra}, {Nicholl}, {Olivares E},
  {Rabinowitz}, {Razza}, {Sollerman}, {Smith}, {Terreran}, {Valenti}, {Gibson},
  \& {Goggia}}]{scs+16}
{Smartt}, S.~J., {Chambers}, K.~C., {Smith}, K.~W., {et~al.} 2016, \mnras, 462,
  4094

\bibitem[{{Sullivan} {et~al.}(2006){Sullivan}, {Le Borgne}, {Pritchet},
  {Hodsman}, {Neill}, {Howell}, {Carlberg}, {Astier}, {Aubourg}, {Balam},
  {Basa}, {Conley}, {Fabbro}, {Fouchez}, {Guy}, {Hook}, {Pain},
  {Palanque-Delabrouille}, {Perrett}, {Regnault}, {Rich}, {Taillet}, {Baumont},
  {Bronder}, {Ellis}, {Filiol}, {Lusset}, {Perlmutter}, {Ripoche}, \&
  {Tao}}]{slp+06}
{Sullivan}, M., {Le Borgne}, D., {Pritchet}, C.~J., {et~al.} 2006, \apj, 648,
  868

\bibitem[{{Sullivan} {et~al.}(2010){Sullivan}, {Conley}, {Howell}, {Neill},
  {Astier}, {Balland}, {Basa}, {Carlberg}, {Fouchez}, {Guy}, {Hardin}, {Hook},
  {Pain}, {Palanque-Delabrouille}, {Perrett}, {Pritchet}, {Regnault}, {Rich},
  {Ruhlmann-Kleider}, {Baumont}, {Hsiao}, {Kronborg}, {Lidman}, {Perlmutter},
  \& {Walker}}]{sch+10}
{Sullivan}, M., {Conley}, A., {Howell}, D.~A., {et~al.} 2010, \mnras, 406, 782

\bibitem[{{Taylor} {et~al.}(2014){Taylor}, {Cinabro}, {Dilday}, {Galbany},
  {Gupta}, {Kessler}, {Marriner}, {Nichol}, {Richmond}, {Schneider}, \&
  {Sollerman}}]{tcd+14}
{Taylor}, M., {Cinabro}, D., {Dilday}, B., {et~al.} 2014, \apj, 792, 135

\bibitem[{{Tonry}(2011)}]{t11}
{Tonry}, J.~L. 2011, \pasp, 123, 58

\bibitem[{{Tully} {et~al.}(2009){Tully}, {Rizzi}, {Shaya}, {Courtois},
  {Makarov}, \& {Jacobs}}]{trs+09}
{Tully}, R.~B., {Rizzi}, L., {Shaya}, E.~J., {et~al.} 2009, \aj, 138, 323

\bibitem[{{Wainscoat} {et~al.}(2016){Wainscoat}, {Chambers}, {Lilly}, {Weryk},
  {Chastel}, {Denneau}, \& {Micheli}}]{wcl+16}
{Wainscoat}, R., {Chambers}, K., {Lilly}, E., {et~al.} 2016, in IAU Symposium,
  Vol. 318, Asteroids: New Observations, New Models, ed. S.~R. {Chesley},
  A.~{Morbidelli}, R.~{Jedicke}, \& D.~{Farnocchia}, 293--298

\bibitem[{{Wyder} {et~al.}(2005){Wyder}, {Treyer}, {Milliard}, {Schiminovich},
  {Arnouts}, {Budav{\'a}ri}, {Barlow}, {Bianchi}, {Byun}, {Donas}, {Forster},
  {Friedman}, {Heckman}, {Jelinsky}, {Lee}, {Madore}, {Malina}, {Martin},
  {Morrissey}, {Neff}, {Rich}, {Siegmund}, {Small}, {Szalay}, \&
  {Welsh}}]{wtm+05}
{Wyder}, T.~K., {Treyer}, M.~A., {Milliard}, B., {et~al.} 2005, \apjl, 619, L15

\bibitem[{{Yasuda} \& {Fukugita}(2010)}]{yf10}
{Yasuda}, N., \& {Fukugita}, M. 2010, \aj, 139, 39

\end{thebibliography}
\end{document}